# Modeling network dynamics: the *lac* operon, a case study

José M. G. Vilar[1], Călin C. Guet[1,2], and Stanislas Leibler[1]

[1]The Rockefeller University, 1230 York Avenue, New York, NY 10021, USA

[2]Dept. of Molecular Biology, Princeton University, Princeton, NJ 08544, USA

Corresponding author: José M. G. Vilar

The Rockefeller University, 1230 York Avenue, Box 34, New York, NY 10021

Phone: (212) 327-7642 / FAX: (212) 327-7640 / e-mail: vilarj@rockefeller.edu

*Abstract*

We use the *lac* operon in *Escherichia coli* as a prototype system to illustrate the current state, applicability, and limitations of modeling the dynamics of cellular networks. We integrate three different levels of description —molecular, cellular, and that of cell population— into a single model, which seems to capture many experimental aspects of the system.

*Introduction*

Modeling has had a long tradition, and a remarkable success, in disciplines such as engineering and physics. In biology, however, the situation has been different. The enormous complexity of living systems and the lack of reliable quantitative information have precluded a similar success. Currently, there is a renewal of interest in modeling of biological systems, largely due to the development of new experimental methods generating vast amounts of data, and to the general accessibility of fast computers capable, at least in principle, to process this data (Endy and Brent 2001, Kitano 2002). It seems that a growing number of biologists believe that the interactions of the molecular components may be understood well enough to reproduce the behavior of the organism, or its parts, either as analytical solutions of mathematical equations or in computer simulations.

Modeling of cellular processes is typically based upon the assumption that interactions between molecular components can be approximated by a network of biochemical reactions in an ideal macroscopic reactor. Although some spatial aspects of cellular processes are taken into account in modeling of certain systems, e.g. early development of *Drosophila melanogaster* (Eldar et al. 2002), it is customary to neglect all the spatial heterogeneity inherent to cellular organization when dealing with genetic or metabolic networks. Then, following standard methods of chemical reaction kinetics, one can obtain a set of ordinary differential equations, which can be solved computationally. This standard modeling approach has been applied to many systems, ranging from a few isolated components to entire cells. In contrast to what this widespread use might indicate, such modeling has many limitations. On the one hand, the cell is not a well-stirred reactor. It is a highly heterogeneous and compartmentalized structure, in which phenomena like molecular crowding or channeling are present (Ellis 2001), and in which the discrete nature of the molecular components cannot be neglected (Kuthan 2001). On the other hand, so few details about the actual *in vivo* processes are known that it is very difficult to proceed without numerous, and often arbitrary, assumptions about the nature of the nonlinearities and the values of the parameters governing the reactions. Understanding these limitations, and ways to overcome them, will become increasingly important in order to fully integrate modeling into experimental biology.

We will illustrate the main issues of modeling using the example of the *lac* operon in *Escherichia coli*. This classical genetic system has been described in many places; for instance, we refer the reader to the lively account by Müller-Hill (Müller-Hill 1996). Here, we concentrate our attention on the elegant experiments of Novick and Weiner (Novick and Weiner 1957). These experiments demonstrated two interesting features of the *lac* regulatory network. First, the induction of the *lac* operon was revealed as an all-or-none phenomenon; i.e., the production of lactose degrading enzymes in a single cell could be viewed as either switched on ("induced") or shut off ("uninduced"). Intermediate levels of enzyme production observed in the cell population are a consequence of the coexistence of these two types of cells (see Figure 1a). Second, the experiments of Novick and Weiner also showed that the state of a single cell (induced or uninduced) could be transmitted through many generations; this provided one of the simplest examples of phenotypic, or epigenetic, inheritance (see Figure 1b). We will argue below that even these two simple features cannot be quantitatively understood using the standard approach for modeling of networks of biochemical reactions. This example will also allow us to explain the different levels, at which biological networks need to be modeled.

*The lac operon*

The *lac* operon consists of a regulatory domain and three genes required for the uptake and catabolism of lactose. A regulatory protein, the LacI repressor, can bind to the operator and prevent the RNA polymerase from transcribing the three genes. Induction of the *lac* operon occurs when the inducer molecule binds to the repressor. As a result, the repressor cannot bind to the operator and transcription proceeds at a given rate. The probability for the inducer to bind to the repressor depends on the inducer concentration inside the cell. The induction process is thus helped by the permease encoded by one of the transcribed genes, which brings inducer into the cell. In this way, if the number of permeases is low, the inducer concentration inside the cell is low and the production of permeases remains low. In contrast, if the number of permeases is high, the inducer concentration is high and the production of permeases remains high.

This heuristic argument is useful for understanding the presence of two phenotypes, but it does not actually explain why the cells remain in a given state, or what makes the cells switch from the uninduced to the induced state. One needs quantitative approaches to understand the dynamics of this process, how the intrinsic randomness of molecular events affects the system, and how induction depends on the molecular aspects of gene regulation.

*Levels of organization and modeling*

Despite its apparent simplicity, the *lac* operon system displays much of the complexity and subtlety inherent to gene regulation. In principle, its detailed modeling should include, among many other cellular processes, transcription, translation, protein assembly, protein degradation, binding of different proteins to DNA, and binding of small molecules to the DNA-binding proteins. In addition, the *lac* operon system is not isolated from the rest of the cell. Induction changes the growth rate of individual cells, which in turn also affects the cell population behavior. For instance, if a gratuitous inducer is used, induction will slow down cell growth. Therefore, extrapolating directly from the molecular level all the way up to the cell population level requires additional information about cellular processes that is not readily available. Moreover, most of the molecular details of the cell are not going to be relevant for the particular process under study. The first step of modeling is, therefore, to identify the relevant levels, their interactions, and the way one level is incorporated into another. Figure 2 illustrates schematically the separation of the *lac* system into molecular, cellular, and population levels.

*Molecular level—* The molecular level explicitly includes the binding of the inducer to the repressor, changes in repressor conformation, binding of the repressor to the operator, binding of the RNA polymerase to the promoter, initiation of transcription, production of mRNA, translation of the message, protein folding, and so forth. Almost all the quantitative aspects of the *in vivo* dynamics of these processes are unknown. The lack of information is typically filled out with assumptions based on parsimony. Fortunately, not all the details are needed. At this level, what seems relevant is the production of permeases expressed as a function of the inducer concentration inside the cell. To obtain theoretically even a rough approximation of this function one would need detailed information about many molecular interactions. Therefore, a more reasonable approach at the present stage of knowledge would be to extract this function directly from the experimental data. Indeed, one can measure the rate of production of β-galactosidase in mutant strains lacking the permease (Herzenberg 1958). In this case, external and internal inducer concentrations are both the same once equilibrium between the medium and the cytoplasm is reached. This relies on the absence of non-specific import or export mechanisms. The other key piece of information is that the production of permeases is, to a good approximation, proportional to the production of β-galactosidase, since both are produced from the same polycistronic mRNA. The results obtained in this way could be used as an estimate for modeling the molecular level of wild type cells.

*Cellular level*—The core of the all-or-none process resides at the cellular level. Some of the permeases produced will eventually go to the membrane and bring more inducer. Novick and Weiner inferred from experiments that only a few percent of the permeases integrate into the membrane and become functional (Novick and Weiner 1959). Recent experiments, however, showed that the majority of the permeases integrate in the membrane (Ito and Akiyama 1991), yet the question of how many are functional has not been addressed. Despite intense studies on the permease (Kaback et al. 2001), its *in vivo* functioning is still a challenging issue, which includes many open questions such as the mechanisms of insertion into the membrane. The simplest assumption for modeling is that the produced permeases are inserted into the membrane and become functional with a constant probability rate. We believe this to be the weakest point of our model. In view of the all-or-none phenomenon, single cell studies on the concentration and the functional state of the permeases would be extremely useful using techniques that are now available (Thompson et al. 2002).

*Population level*— Induction of the *lac* operon changes the growth rate of the cells. When lactose is the sole carbon source, induction allows cells to grow. For gratuitous inducers, like the one used in Novick and Weiner experiments, the situation is just the opposite: induction slows down the growth rate. This slowing down seems to be connected with the number of permeases in the membrane (Koch 1983). At this level, it seems adequate to use a standard two-species population dynamics model. The growth rates for induced and uninduced cells are known from the experiments. The cellular level is integrated into the population level by considering the induced-uninduced switching rates. These rates can be obtained by modeling at the cellular level by computing the probability for an uninduced cell to become induced and for an induced cell to become uninduced.

*The Model*

The preceding discussion seems to indicate that three variables are relevant for the description of the functioning of the *lac* system. These are the concentrations of non-functional permease ($Y$), of functional permease ($Y_f$), and of inducer inside the cell ($I$). Another variable that we need to incorporate explicitly in the model is the concentration of $\beta$-galactosidase ($Z$), which is the quantity measured in the experiments.

Now, we are ready to model the dynamics of the induction process by writing down the phenomenological dynamical equations for these variables:

$$\frac{dY}{dt} = f_1(I) - a_1 Y$$

$$\frac{dY_f}{dt} = b_1 Y - a_2 Y_f$$

$$\frac{dI}{dt} = [f_2(I_{ex}) - f_3(I)]Y_f + b_2 I_{ex} - a_3 I$$

$$\frac{dZ}{dt} = g\, f_1(I) - a_3 Z$$

Here, $I_{ex}$ is the external inducer concentration; $g$, $b_1$, $b_2$, $a_1$, $a_2$, and $a_3$ are constants; and $f_1$, $f_2$, and $f_3$ are functions of their respective arguments. The molecular level description enters the equations through the specific form of $f_1$, $f_2$, and $f_3$. $f_1(I)$ is the production rate of permeases as a function of the internal inducer concentration. As explained above, it can be obtained from experiments. It behaves like a quadratic polynomial for low inducer concentrations ($f_1(I) \cong c_1 + c_2 I + c_3 I^2$, with $c_1$, $c_2$, and $c_3$ constants) and increases monotonically until it saturates for high concentrations. The functions $f_2(I_{ex})$ and $f_3(I)$ account for the inducer transport by the permease in and out of the cell and are assumed to depend hyperbolically on their argument.

With only these four equations one can explain the fact that there are inducer concentrations, $I_{ex}$, for which the cells remain induced, if they were previously induced, or uninduced, if they were uninduced. In mathematical terms, this happens because the equations have two stable solutions for such values of $I_{ex}$ and the system thereby exhibits "hysteresis". Thus, the standard modeling approach can apparently explain the existence of the so-called maintenance concentration.

There are many variations of this simple model. The first one, proposed already by Novick and Weiner, was even simpler and explained to some extent the main features observed in the experiments (Novick and Weiner 1957, Cohn and Horibata 1959a,b). In fact, subsequent, much more complex models, based on the standard biochemical reaction kinetics approach, did not provide any substantial additional insight. They basically showed that the observed behavior is also compatible with more intricate kinetics (Chung and Stephanopoulos 1996).

To fully understand the all-or-none phenomena the standard approach is, however, not enough. One needs to take into account stochastic events to explain why, at some point, just by chance, a cell becomes induced. The classical approach is unable to explain the switch from the uninduced to the induced state. Fortunately, it is possible to write down a stochastic counterpart of the previous equations. This is done by transforming the different rates (production, degradation, etc.) into probability transition rates and concentrations into numbers of molecules per cell. Then, one can simulate the dynamical behavior of the four random variables governed by such stochastic equations on a computer (Gillespie 1977).

Figure 3a shows representative time courses of the $\beta$-galactosidase content obtained from such computer simulations for cells placed under sub-optimal induction conditions. At the single cell level, there is a fast switch from the non-induced to the induced state. The time at which this transition happens is a result of the intrinsic stochastic nature of biochemical reactions and strongly varies from cell to cell (see e.g. yellow, green, and blue lines in Figure 3a). In contrast, the cell average exhibits a smooth behavior. In this case, the behavior of the single cell and the behavior of the cell average are thus completely different. As a consequence, classical reaction kinetics cannot be used and has to be replaced by a stochastic approach. This type of approach started to be applied in the 1940s (Delbrück 1940) and was already well established in the late 1950s (Montroll and Shuler 1958). Only recently, however, there has been a renewed widespread effort to understand the role of stochasticity in cellular processes (Rao et al. 2002).

One should stress that even the stochastic approach is still unable to fully explain the experiments. In the simulations, all the cells eventually become induced. In the experiments, the production of $\beta$-galactosidase for sub-optimal inducer concentrations seems not to saturate at the maximum value, which is an indication of the coexistence of the induced and uninduced cells. As explained before, the reason for this is that the induced and uninduced cells grow at a different rate. Therefore, we have to consider the dynamics of the cell population. Only when this is taken properly into account, the simulations are in agreement with experiments, as shown by the dashed line in Figure 3a.

The fact that fluctuations make cells switch from the uninduced to the induced state forces us to reconsider whether there really exists a maintenance concentration in the model. Is there a range of inducer concentrations for which the cells do not switch at an appreciable rate from one state to another? Figure 3b shows the single cell behavior for cells that were previously induced or uninduced at the expected maintenance concentration. Indeed, in the simulations we performed for 1000 cells, we recorded no single switching from one state to another: for realistic values of probability rates such switching events would be too rare to be observed. The stochastic model seems to be thus compatible with the existence of the maintenance concentration.

So far, we have pointed out just a few of the many limitations of the standard modeling approach and how to overcome them. Considering stochastic and population effects greatly increases the complexity of modeling. In general, whether or not we should consider all of these effects depends not only on the given system but also on the particular conditions. For instance, an approach taking into account all three levels of description is not needed when the *lac* operon is induced at high inducer concentrations. In this case, the single cell picture, the average over independent cells, and the population average all give very similar results, as can be seen in Figure 3c. Therefore, it should be possible here to use standard kinetic equations and avoid most of the hassle encountered beyond the standard approach. The main problem, however, is that there is no general *a priori* method to tell whether or not the standard modeling approach would be sufficient to describe the given system.

In Figure 3d we compare experimental and simulation results. There are some differences: the rise in $\beta$-galactosidase activity is faster in the experiments than in the simulations. In addition, coming back to Figure 3b, one can see that there is a small drop in $\beta$-galactosidase content when cells are transferred from high to maintenance inducer concentrations. This drop is not present in the experiments (see Figure 3 in Novick and Weiner 1957). One cannot infer from the model whether these differences are a matter of details or of a more fundamental aspect of the *lac* system. The addition of more molecular details into a model (Carrier and Keasling 1999) does not necessarily lead to better agreement with the experimental observations. The *lac* operon example clearly illustrates the complexity of modeling even the simplest networks.

*Evolutionary and physiological levels*

The type of models and experiments that we have discussed can provide valuable information about the mechanistic structure of the *lac* operon. But, to really understand the functioning and underlying logics of cellular networks, one needs to consider them in their natural environment. Only then is it possible to relate the network structure to the function it has acquired through evolution (Savageau 1977). In the case of the *lac* operon of *E. coli*, induction usually takes place in the mammalian digestive tract under anaerobic conditions (Savageau 1983) and the inducer is allolactose, a metabolic product of lactose, rather than gratuitous inducers, such as IPTG or TMG. In addition there can be other factors that can affect the induction process itself. For instance, recent genetic studies have uncovered a novel set of sugar efflux pumps in *E. coli* that surprisingly can pump lactose outside of the cell (Liu et al. 1999a,b)! The physiological role of these pumps has just started to be investigated.

*Discussion*

The example of the *lac* operon switch has been used here to illustrate the current state, applicability, and limitations of modeling of cellular processes. We have not tried to expose all the potential that modeling possesses: there are now many published reviews advertising this aspect. Rather, we have tried to use one of the simplest and best-studied examples to show the intricacy of modeling biological networks. Here are some ideas that we would like to emphasize:

- Standardized modeling methods cannot be applied "automatically" even in a case as simple as the one we have described. One needs first to identify the relevant variables, adequate approximations, etc. Adding more equations to include more details of interactions does not usually help. If more molecular details are considered, one can easily end up with huge sets of equations, but unless the relevant elements are identified, the model will remain useless. The problem is thus more conceptual than technical. In the case we have discussed, a four-equation model is able to explain the main results of the experiments of Novick and Weiner, provided that fluctuations and population effects, which are usually overlooked, are taken into account.

- One of the main reasons for the success of models in matching the experimental results is that the experiments are kept under constant conditions and only a few variables are changed. This allows the use of effective (fitting) parameters in the equations.

- Networks are neither isolated in space nor in time. They form part of a unity that has been shaped through evolution. It is important not to disregard *a priori* any of the many complementary levels of description: molecular, cellular, physiological, population, intra-population or evolutionary.

In our opinion, because of these and similar reasons, productive modeling of biological systems, even in the "post-genomic era", will still rely more on good intuition and skills of quantitative biologists than on the sheer power of computers.

Kaback H.Rr, M.Sahin-Toth and A.B.Weinglass. 2001. The kamikaze approach to membrane transport. *Nat. Rev. Mol. Cell. Bio.* 2: 610-620.

Kitano H. 2002. Computational systems biology. *Nature* 420: 206-210.

Koch A.L. 1983. The protein burden of Lac operon products. *J. Mol. Evol.* 19: 455-462

Kuthan H. 2001. Self-organisation and orderly processes by individual protein complexes in the bacterial cell. *Prog. Bioph. Mol. Bio.* 75: 1-17.

Liu J.Y., P.F.Miller, M.Gosink, and E.R.Olson. 1999. The identification of a new family of sugar efflux pumps in Escherichia coli. *Mol. Microbiol.* 31:1845-1851.

Liu J.Y., P.F.Miller, J.Willard and E.R.Olson. 1999. Functional and biochemical characterization of Escherichia coli sugar efflux transporters. *J. Biol. Chem.* 274: 22977-22984.

Montroll E.W. and K.E. Shuler. 1958. The application of the theory of stochastic processes to chemical reactions. *Adv. Chem. Phys.* 1: 361-399.
15

**Figure Captions**

**Figure 1**

**(a)** *All-or-none phenomenon.* For low inducer concentrations, the enzyme ($\beta$-galactosidase) content of the population increases continuously in time. This increase is proportional to the number of induced cells, represented here by full ellipses. Empty ellipses correspond to uninduced cells.

**(b)** *Maintenance concentration effect*. When induced cells at high inducer concentration are transferred to the maintenance concentration, they and their progeny will remain induced. Similarly, when uninduced cells at low inducer concentration are transferred to the maintenance concentration, they and their progeny will remain uninduced.

**Figure 2**

**(a)** *Molecular level*. The three genes *lacZ*, *lacY*, and *lacA* are cotranscribed as a polycistronic message from a single promoter P1. The gene *lacZ* encodes for the $\beta$-galactosidase, which can either break down lactose into $\beta$-D-galactose and D-glucose or catalyze the conversion of lactose into allolactose, the actual inducer. The product of *lacY* is the $\beta$-galactoside permease, which is in charge of the uptake of lactose inside the cell. The role of *LacA* is not yet fully understood since its product, a galactoside acetyltransferase, is not required for lactose metabolism (Wang et al.2002). The *lac* repressor is encoded by *lacI*, which is immediately upstream the operon. Binding of the repressor to the main operator site O1 prevents transcription. Repression is greatly enhanced by the additional simultaneous binding of the repressor to one of the auxiliary operator sites O2 and O3. The inducer inactivates the repressor by binding to it and changing its conformation. Additionally, the CAP-cAMP complex must bind to the activator site, AS, for significant transcription.

**(b)** *Cellular level*. Some gratuitous inducers, such as TMG, also use the lactose permease to enter the cell; but in contrast to lactose, they bind themselves to the repressor and are not metabolized by the cell. In this case, it is possible to study the dynamics of induction by considering as variables only the internal inducer concentration, the non-functional permeases, and the functional permeases.

**(c)** *Population level*. Coexistence of two types of networks in the *lac* operon is a population effect. Uninduced cells (empty circles) have some probability to become induced (full circles). If uninduced cells grow faster both types of cells could coexist; if not, the entire population will eventually be induced.

**Figure 3**

**(a)** *Single cell, cell average, and population behavior*. The thin (yellow, green, and blue) color lines correspond to representative time courses of $\beta$-galactosidase content obtained from computer simulations for single cells at 7 μM TMG. The observed differences in switching times from non-induced to induced states result from the stochastic behavior of the model. The thick continuous black line is the average over 2000 cells. The dashed black line is the population $\beta$-galactosidase content. To obtain the population results we have considered in the simulations that induced cells grow slower than uninduced ones.

**(b)** *Maintenance concentration*. Representative time courses of the $\beta$-galactosidase content obtained for induced (top) and uninduced (bottom) cells transferred to the maintenance concentration (5 µM TMG) at time 0. Note the semi-logarithmic scale.

**(c)** Same as in Figure 3a but now for cells at 500 µM TMG. In this case the cell average and population $\beta$-galactosidase content are indistinguishable.

**(d)** *Simulation vs. experiments*. Induction for 500 µM TMG at the population level. The black line is the same as in Figure 3c. Red dots represent experimental values obtained by Novick and Weiner (Novick and Weiner 1957). The dashed line is the same as the black line but shifted to the left 0.33 generations. It illustrates that the main differences between simulations and experimental results come from the early stages of induction.

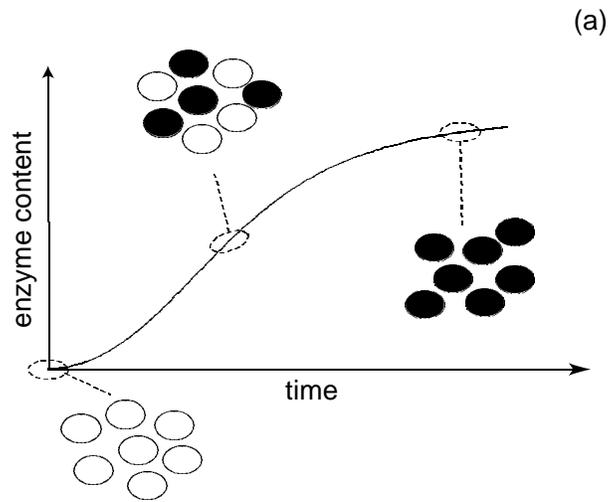
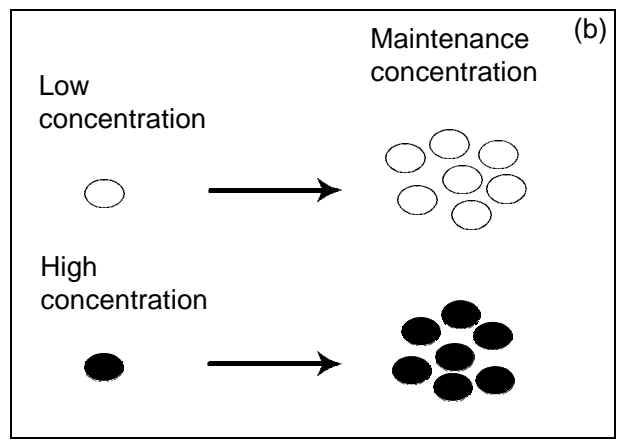

FIGURE 1

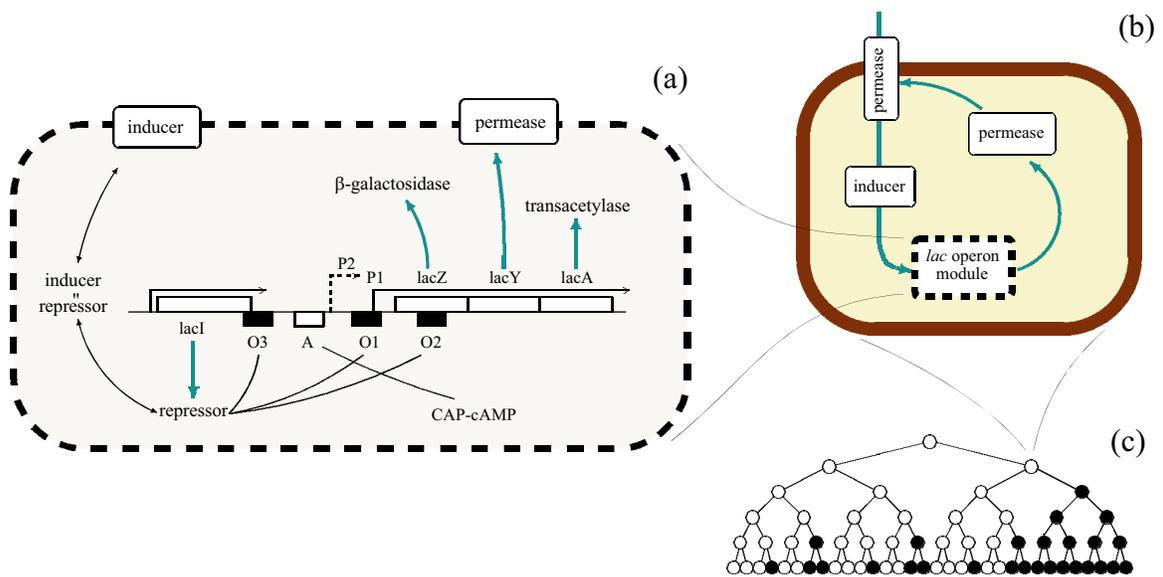

FIGURE 2

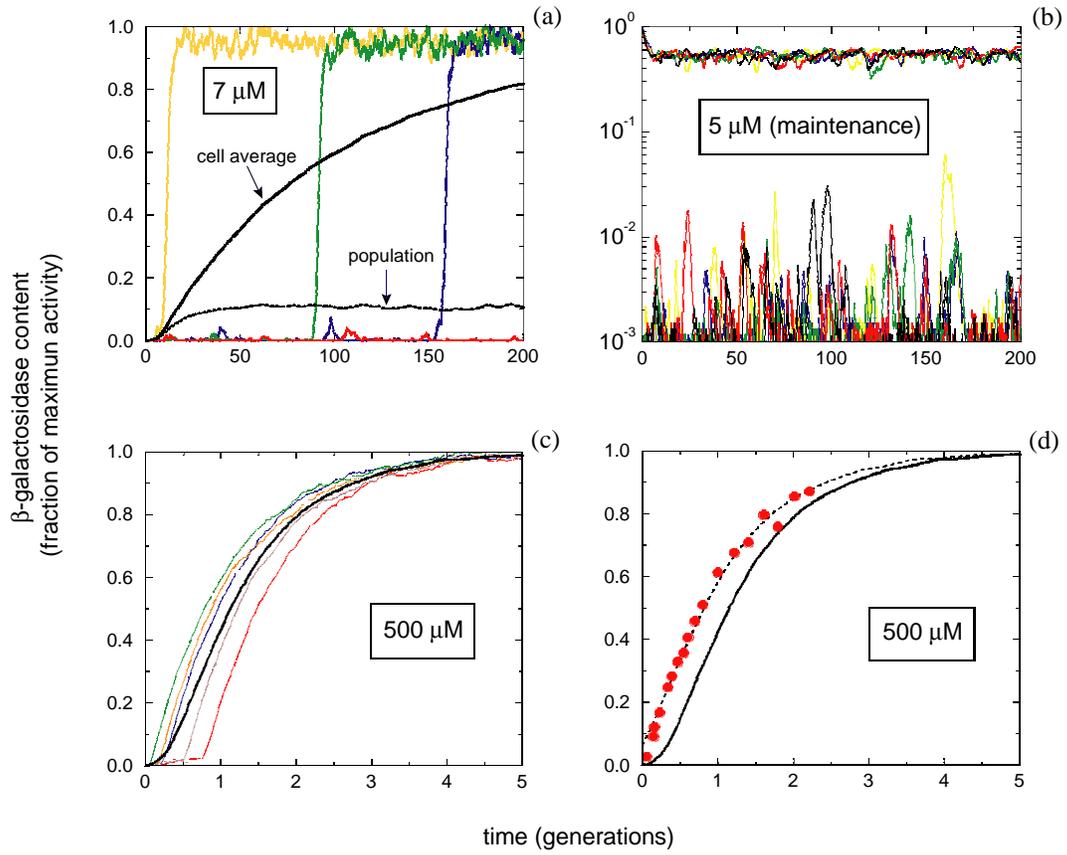

FIGURE 3